\documentclass[12pt]{amsart}
\usepackage{amsmath,amssymb,amscd,latexsym,graphicx,epsfig,color}
\begin{document}

\markboth{John Conway and Simon Kochen}{Thou Shalt Not Clone One Bit!}

\begin{flushleft}
{\bf THOU SHALT NOT CLONE ONE BIT!}
\end{flushleft}


\begin{quote}
\noindent {\bf John Conway$^1$} {\bf and Simon Kochen}\footnote{Princeton
University, Department of Mathematics, Princeton, NJ 08544-1000: \\
\indent \ \ jhorcon@yahoo.com; kochen@math.princeton.edu}
\end{quote}

\setcounter{footnote}{0}
\renewcommand{\thefootnote}{1}

\vskip 1.0cm 

\begin{flushleft}
{\bf 1.  INTRODUCTION}
\end{flushleft}

\vskip .5cm

\parskip .4cm
Several authors have found results that are
collectively known as ``The No-Cloning Theorem,"
to the effect that it is not possible to put
one elementary particle into a state that is
an exact copy of that of an arbitrarily given one,
without disturbing the state of that given one.(See e.g. [1]).
On the other hand, in our paper [2] on ``The Free Will Theorem,"
we made use of the fact that it is possible for ``twinned"
spin 1 particles to exist, meaning just that they will
give the same answer to the question of whether the
component of spin in any direction is zero or not.

In this paper we prove the following:
\begin{list}{1)}{}
\item The No-Triplets Theorem.
\end{list}
{\it It is impossible for there to exist three spin 1 particles
any two of which are twinned in the above sense.}

This rather trivially implies our own
\begin{list}{2)}{}
\item Strong No Cloning Theorem for Spin 1 Particles
\end{list}
{\it It is impossible to ``weakly clone" a spin 1 particle ---
that is to say, without disturbing its state, to produce
the above kind of ``twin" for it.}

For if this were possible, we could produce a set of
triplets by cloning one of a pair of twins. There is in fact
a strengthening of the No-Triplets Theorem, namely:
\begin{list}{3)}{}
\item The ``No-Twins-and-a-Bit" Theorem.
\end{list}
{\it It is impossible for there to exist three spin 1
particles of which two are twinned, while the third will
give the same answer that those two do to a spin-zero
measurement in just one fixed direction.}

Finally, this yields the theorem of our title:
\begin{list}{4)}{}
\item Thou Shalt not Clone one Bit.
\end{list}
Namely, {\it that without disturbing the state
of a given spin 1 particle, one cannot even produce
another one that must give the same answer to a spin-zero
measurement in just one fixed direction, (which provides
just one information-bit - 0 for spin zero, and 1 otherwise).}

For by doing this to one of a twinned pair, one
would produce twins-and-a-bit.

\setcounter{footnote}{0}
\renewcommand{\thefootnote}{2{}}
We remark that since the binary digit produced by a
spin-zero measurement has probability 2/3 of being 1 and
1/3 of being 0, it is not quite one bit, for in
Shannon's sense its information content is only
$$
-(2/3)\log_2(2/3)-(1/3)\log_2(1/3)  =  0.9182958336 \dots < 45/49 \text{ bit}^{\;2} 
\footnotetext{We were tempted to title our paper ``Thou shalt not clone 45/49
of a bit," remembering Charles Babbage's alleged response to the couplet
\begin{quote}
    ``Every minute dies a man,
     every minute one is born"
\end{quote}
from Tennyson's poem ``The Vision of Sin."
Babbage supposedly wrote to the author that this should be corrected to read
\begin{quote}
    ``Every minute dies a man,
     and one and one-sixteenth is born"
\end{quote}
since the original ``would tend to keep the sum total of the world's
population in a state of perpetual equipoise, whereas it is a
well-known fact that the sum total is constantly on the increase."

He went on to say that ``I may add that the exact figures are 1.067, 
but something must, of course, be conceded to the laws of metre."}
$$

\begin{flushleft}
{\bf 2. THE PROOF}
\end{flushleft}

By what has already been said, it suffices to prove
assertion 3).  We shall suppose to the contrary that
spin 1 particles A, B and C exist, that A and B are
twinned, while a spin-zero measurement of C in the
$z$-direction necessarily produces the same answer as the
similar measurement for A or B in that direction.

We let  $u$, $v$  and  $w$  be unit vectors, and  $i$, $j$, $k$  be
chosen from $0,1$.  Then we write 
$Pr(u \rightarrow i, v \rightarrow j, w \rightarrow k)$
for the probability that spin-zero measurements for our three
particles in the respective directions  $u$, $v$, $w$  will produce
the bits $i$, $j$, $k$.  We use the ``$1,0,1$  property" that 
$i$, $j$, $k$ must necessarily
be $1,0,1$ in some order (since for a spin 1 particle the
squared spin components $s_x^2$, $s_y^2$, $s_z^2$ commute and have
sum 2).

Then the corresponding probability, $Pr(v \rightarrow i, w \rightarrow  j)$,
for just our first two particles, since they are twinned,
equals the probability that successive measurements of a
single particle in directions  $u$ and $v$ will give the
answers  $i$ and $j$, and quantum mechanics gives
$$
\begin{aligned}
   Pr(u \rightarrow 0, v \rightarrow  0)& 
     =  (1/3)(u.v)^2  \hskip 2.0in \qquad  (\star)\\
   Pr(u \rightarrow 0, v \rightarrow 1)& 
     =  (1/3)( 1 - (u.v)^2 ) = Pr(u \rightarrow 1, v \rightarrow 0) \;
\qquad(\star\star)
\end{aligned}
$$
(for a proof, see Endnote 7 of [2]).

Fig, 1: The seven vectors of the proof, joined when orthogonal.

\bigskip

\begin{figure}[!h]
\centering
\includegraphics{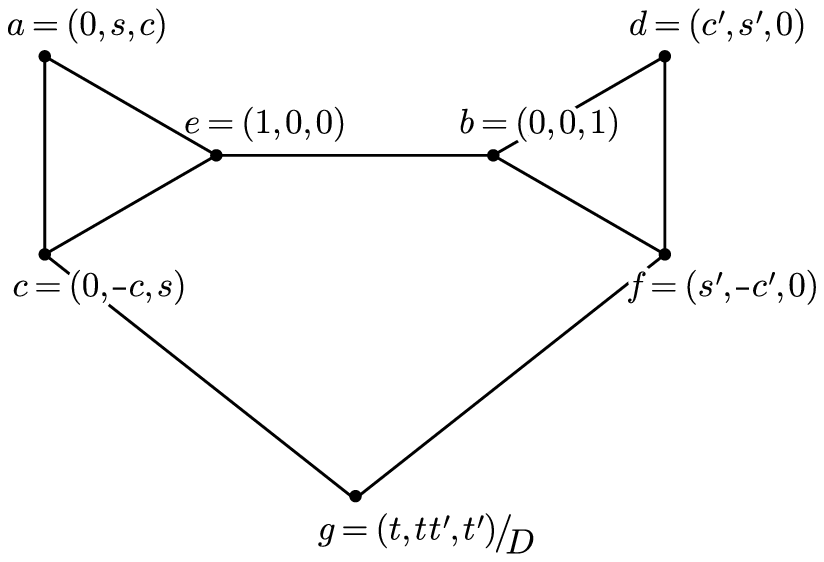}
\end{figure}

\vskip .5cm

We now have, for the vectors $a,b,c,d,e,f,g$  of Figure 1, in which
$s,c,t$  and  $s',c',t'$  denote the sines, cosines, and tangents of
the angles from  $a$ to $b$  and  $d$ to $e$  respectively, while
$D^2 = t^2 + (t')^2 + (tt')^2$:

\begin{tabular}{lr}
$Pr(a \rightarrow 0, d \rightarrow 0, g \rightarrow 1)$&\\
$\qquad = Pr(a \rightarrow 0, d \rightarrow 0) - Pr(a \rightarrow 0, d
\rightarrow 0, g \rightarrow 0)$ &        (1) \ \\

$\qquad = (1/3)(ss')^2 - Pr(e \rightarrow 1, b \rightarrow  1, 
g \rightarrow 0)$  &               (2) \ \\
$\qquad = (1/3)(ss')^2 - Pr(e \rightarrow 1, g \rightarrow 0) + 
    Pr(e \rightarrow 1, b \rightarrow 0, g \rightarrow 0)$& (3) \ \\
$\qquad = (1/3)(ss')^2 - ( 1/3 - t^2/3D^2 ) + 
     Pr(b \rightarrow 0, g \rightarrow 0)$  &    (4) \ \\
$\qquad = (1/3)(ss')^2 - 1/3 + t^2/(3D^2)  +  t'^2/(3D^2)$ &     (5) \ \\ 
$\qquad = -(1/3)(ss'/D)^2 $&       \ \ (6).
\end{tabular}

The fact that this probability is strictly negative (when $b$ and $e$
are in distinct directions from $a$ and $d$  respectively) proves our theorem.

Finally, we explain the successive equalities:

(1) and (3) apply the addition law of probabilities.
(2) uses (*) and the fact that  $a \to 0, d \to 0, g \to 0$
happens if and only if  $e \to 1, b \to 1, g \to 0$, as easily follows
from the $1,0,1$ property.
(4) similarly uses ($\star\star$) and the fact that  $b \to 0, g \to 0$ 
entails $e \to 1$ by the $1,0,1$ property.
Finally, (5) is another application of ($\star$), and (6) is 
some trigonometrical manipulation.

\noindent{\bf REFERENCES}

[1] W. K. Wootters and W. H. Zurek, A Single Quantum Cannot be Cloned,
    Nature 299 (1982), 802-803.

[2] J. Conway and S. Kochen, The Free Will Theorem, Found. Phys. 36
    (2006), 1441-1473.

\end{document}